# A physicist's view of DNA

Alireza Mashaghi & Allard Katan

Cees Dekker lab, Bionanoscience Department, Kavli Institute of Nanoscience, Delft University of Technology.

Nucleic acids, like DNA and RNA, are molecules that are present in any life form. Their most notable function is to encode biological information. Why then would a physicist be interested in these molecules? As we will see, DNA is an interesting molecular tool for physicists to test and explore physical laws and theories, like the ergodic theorem[1], the theory of elasticity[2] and information theory. DNA also has unique material properties, which attract material scientists, nanotechnologists and engineers. Among interesting developments in this field are DNA-based hybrid materials and DNA origami[3-6].

## DNA, the regular aperiodic molecule

In his 1945 book "What is life?" Schrodinger[7] postulated that for a molecular system to carry information, it has to be regular but aperiodic. In his view, irregular amorphous materials are too chaotic to be able to practically carry information, but full periodicity reduces the capacity to encode information. Was Schrodinger right?
In principle not, since we now know there are ways to encode information even in irregular materials. But in the case of DNA, Schrodinger's prediction was crucial to the discovery of the structure. Not only did he inspire theoretical physicist Francis Crick and biochemist James Watson to look in the right direction, the fingerprint of a regular structure that was visible in X-ray diffraction data was instrumental in unraveling the now famous double helix structure of DNA[8]. The DNA molecule is indeed structurally periodic, but aperiodic at the sequence level (Figure 1A).

## The structure of DNA

DNA is an organic polymer of four different monomers. Each monomer is composed of a phosphate group, a single-ring sugar and one of four bases: A, T, C and G. The bases are planar single- or double-ring aromatic compounds. The monomers can form bonds between the sugar and the phosphate and form into a long polymer or strand of DNA. Such a single strand is irregular and if it consists of N monomers, there can be $4^N$ different combinations of bases, so the molecule carries $2^{2N}$ bits of information. The bases can also interact with each other through the formation of hydrogen bonds. Specifically, A pairs with T and G pairs with C. The energy gain of base-pairing is not the same for these two pairs, since A and T only share 2 hydrogen bonds, while C and T share 3. An entire strand of DNA can form base-pairs with a complementary strand to make double-stranded DNA. The chirality of the sugars and the optimization of hydrogen bond angles leads to the formation of a regular helical structure with 10.4 base pairs per turn. Since the sequence of the complementary strand is entirely determined by the template strand, double stranded DNA does not carry more information than single stranded, even though it uses twice the amount of material. In spite of this, almost all organisms encode their genetic information on double-stranded DNA.

## Twisting DNA beyond the helix: supercoiling

To read the base sequence, proteins use the same hydrogen bonding sites as used for DNA pairing. Therefore reading requires breaking of base-pairing, so-called melting. Melting unwinds the helix and allows the information to be read. When a protein walks along the DNA and locally melts it, it displaces the helical twist. Just like a normal rope that is twisted, DNA that is wound up will at some point collapse into supercoils, or plectonemes as they are called.
Recently, researchers in the Cees Dekker lab have managed to visualize such supercoils[9]. To do this, they attached a DNA molecule to a magnetic bead on one side, and a glass slide on the other side. Using external magnets, they rotated the DNA, mimicking the action of an unwinding protein and forming supercoils. By fluorescently labeling the DNA, and pulling the magnetic bead to the side, they could see where the supercoils were located, and follow how they are created and annihilated, or moved spontaneously along the DNA.

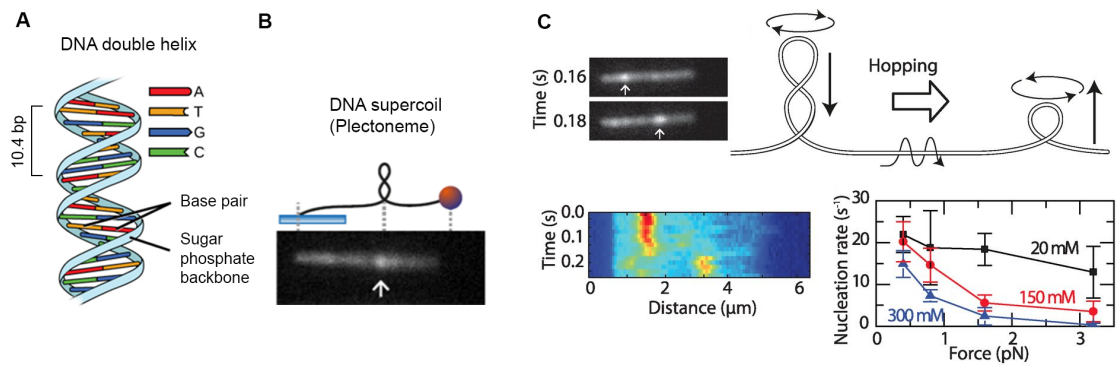

 Figure 1. (A) Structure of DNA double helix (B) DNA supercoiling visualized with fluorescence and magnetic tweezers. Schematic illustrating how higher fluorescence intensity reveals the position of a plectoneme. (C) plectoneme hopping from one position to another and DNA. Analysis of different experimental conditions shows that the rate of formation of new plectonemes depends on salt concentration and pulling force.

## DNA and theory of elasticity

The single molecule measurements of force and extension of DNA have provided the most rigorous test to date of theories of entropic elasticity[2]. When free in solution, a flexible polymer randomly coils, wraps around itself and adopts various conformations, leading to an average end-to-end distance much shorter than its contour length. In contrast, when a linear polymer is stretched from its ends, fewer conformations are available to the molecule. Pulling the molecule into a more extended chain is thus entropically unfavorable, resulting in an entropic resistance against pulling.

Various approaches have been used to stretch DNA molecules including magnets, fluid flow and later optical traps. The entropic force–extension behavior of double stranded DNA agreed well with the worm like chain model. Forces exerted by molecular motors found in cells, can stretch double stranded DNA to nearly its contour length.

In addition to its use in reading the information, the base pairing also provides mechanical strength to the DNA molecule. The persistence length of double stranded DNA is ~50 nm, at least a factor of five larger than the dimension of the molecular machines that read the information on the DNA. Single stranded DNA is much more floppy, it's persistence length is only 1 nm.

## A Physical way to read DNA

Reading the information encoded in DNA is extremely important for biomedical sciences. Much effort has been put in to develop technology for sequencing DNA, which is the accurate reading of the order of bases in a DNA strand. A new technique for sequencing, that is still in early stages of development, is based on electrical current measurements. Simulations and theoretical studies suggest that it is possible to sequence a DNA molecule during its translocation through a nanopore, a few nm wide hole in a solid state membrane. For instance it is proposed[10] that by passing single DNA molecules through a nano-gap between two electrodes and measuring the electron tunneling, one could read the DNA sequence. Electron tunneling through a nano-gap is affected differently by the presence of different DNA bases in the gap. Besides the future promise of sequencing, solid state nanopore measurements are already being used to determine the presence of proteins on single DNA molecules, or detect secondary structure[11,12].

## DNA origami: using DNA as a nanoscopic building material

One important difference between DNA and a typical polymer is that the latter is synthesized via chemical means, and has a distribution of molecular weights. The synthesis of DNA by living cells or isolated cell components is performed such that (nearly) all synthesized molecules are exactly the same, not only with the same number of each monomer in the polymer, but also with the same order of the monomers. This precise manufacturing, along with the known base pairing, has led scientists to use DNA as a building material for nanostructures called DNA origami (Figure 2). Using keenly selected short 'staple' strands, a long single-stranded DNA molecule is wrapped up into a desired shape. The nanometer scale precision attainable with these structures is nearly impossible with conventional methods. Though the first origami structures - smiley faces 50 nm in size – were not very useful, proof-of-concept applications in drug delivery have already been demonstrated[13] and more and more complex 3D structures are being built out of DNA. In our lab, DNA origami is used as a tool to enhance the functionality of nanopore and AFM measurements.

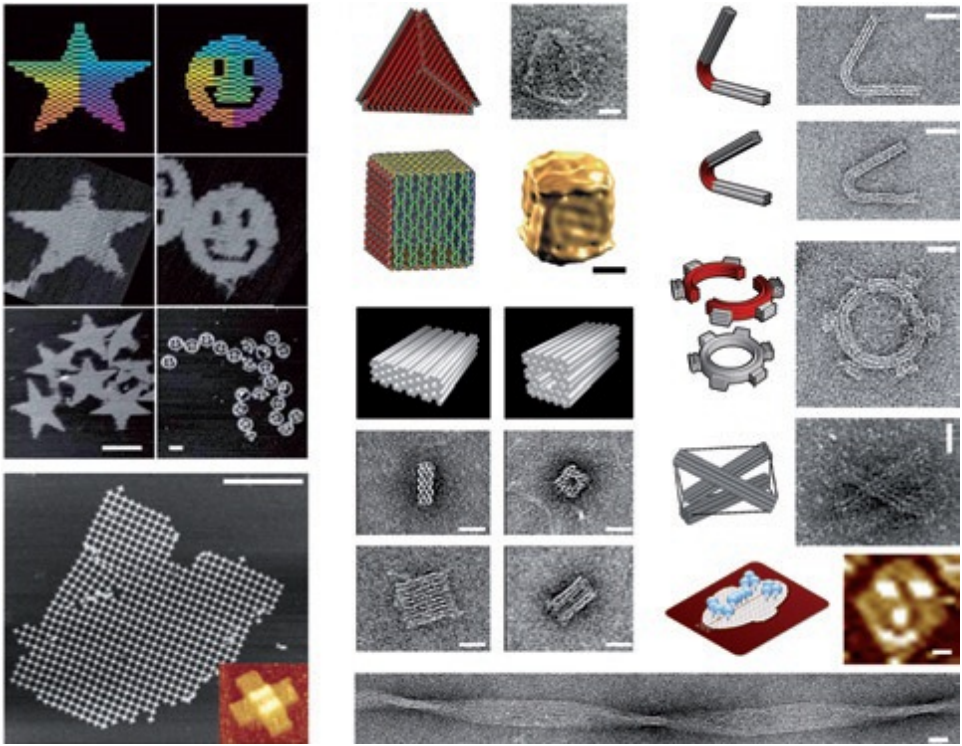

**Figure 2:** DNA origami objects. Top left: models and AFM images of two-dimensional DNA origami structures. Bottom left: a crystal made of DNA tiles. Center and right: models and electron micrographs of various 3-dimensional DNA origamis. Bottom right: model and AFM image of proteins bound to DNA origami. Figure reproduced from Castro et al.[14]

**Ways to visualize a single DNA molecule**

A single DNA molecule is only 2 nm in size, too thin to be visible under a light microscope. Fluorescently labeling the DNA, like in the experiment of Figure 1B, can work around this. There are many uses of fluorescent imaging of DNA, but still it has some drawbacks. Resolution is limited to that of the optical microscope, unlabeled objects are invisible, and biological function can be compromised by the labeling. Electron microscopes can visualize stained DNA, but they can only operate under unnatural conditions.

The Atomic Force Microscope (AFM) is a mechanical microscope, it forms an image by gently 'feeling' the surface of objects. The AFM can resolve single DNA molecules even in a dense environment, can operate in aqueous solution and requires no labeling. The latest developments in AFM technology have pushed the time resolution from a few minutes to tenths of seconds. Such a high-speed AFM – one of the first in Europe- is available in our laboratory. This allows us to make real-time 'molecular movies' of the interactions between DNA and the proteins that maintain it (Figure 3).

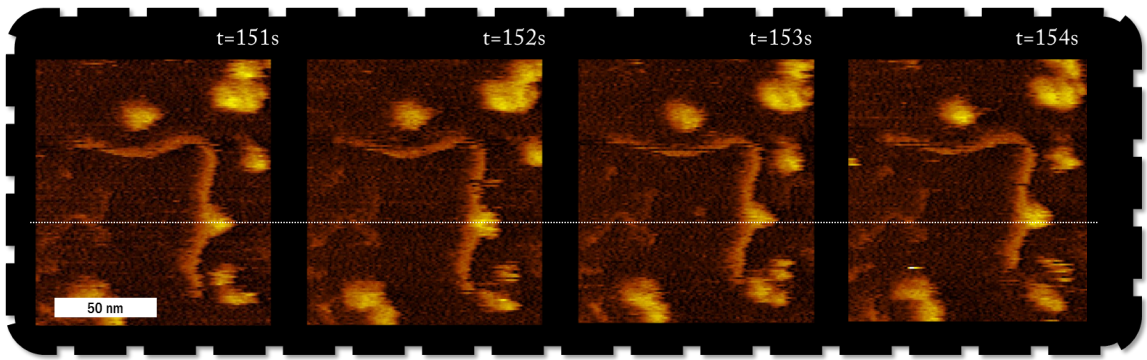

**Figure 3.** A protein complex (tetrasome) hopping to a new position on a DNA molecule. The hopping distance is only 4 nm. Images by Allard Katan, TU Delft

### Exciting opportunities in DNA physics and engineering

More than a century has passed since the discovery of DNA in late 19$^{th}$ century by Friedrich Miescher. Still there are mysteries around the physical properties of DNA. How does DNA respond to force and negative torque? DNA in cellular nuclei is highly constrained and confined. How does confinement affect the structure and dynamics of DNA?

Among most important opportunities in DNA nanotechnology is to fabricate nanomachines with useful functions. We have recently witnessed promising developments in this direction, such as programmable DNA based motors that move on linear tracks[15] and DNA nanorobots that kill cancer cells[13].

DNA may be the quintessential molecule of biology, but as you have seen it is also a rich source for physicists, both as a subject in itself and as a tool to reach new capabilities.